\documentclass[reprint,aps,prc,twocolumn,floatfix,10pt]{revtex4-1}
\usepackage{graphicx,amsmath,bm,color}

\usepackage{dcolumn}
\usepackage{amssymb}
\usepackage{xcolor}
\usepackage{fix-cm}
\usepackage{mathptmx} 
\usepackage[T1]{fontenc}
\usepackage[colorlinks,allcolors=blue]{hyperref}

\begin{document}
\title{Fermions with long and finite range interactions on a quantum ring}
\author{Alexander W.\ Bray}
\affiliation{Department of Theoretical Physics, Research School of Physics, The Australian National University, Canberra ACT 2601, Australia}
\author{C\'edric Simenel}\email{cedric.simenel@anu.edu.au}
\affiliation{Department of Theoretical Physics and Department of Nuclear Physics, Research School of Physics, The Australian National University, Canberra ACT 2601, Australia}
\date{\today}
\begin{abstract}
\begin{description}
\item[Background] 
Idealised systems are commonly used in nuclear physics and condensed matter. For instance, the construction of nuclear  energy density functionals involves properties of infinite matter, while neutron drops are used to test nuclear interactions and approximations to the nuclear many-body problem. In condensed matter, quantum rings are also used to study properties of electron systems. 
\item[Purpose] 
To investigate the possibility to use  quantum rings with systems of nucleons including many-body correlations. 
\item[Methods] 
A quantum ring model of a finite number of same spin fermions is developed. Several attractive and repulsive interactions with finite and infinite ranges are considered. 
Quantum Monte Carlo calculations are used to provide exact ground-state energies. Comparisons with analytical Hartree-Fock solutions are used to get an insight into the role of correlations. 
\item[Results] 
Hartree-Fock results with no breaking of space translational symmetry are able to describe many systems. However, additional spatial correlations are required in the case of dense systems with a strong short-range repulsion, or with attractive interactions in large rings. 
\item[Conclusions] 
Self-bound systems of fermions with spatial correlations produced by basic features of the nuclear interactions can be described on a quantum ring, encouraging  applications with realistic interactions, as well as investigations with higher dimensional geometries such as spherium. 

\end{description}
\end{abstract}
\maketitle
\section{Introduction}

The complexity of the nuclear many-body problem often requires the use of simplified models for idealised systems. Examples include infinite nuclear matter (see, e.g.,  \cite{oertel2017,leonhardt2020}),  nuclear slabs \cite{bonche1976,rios2011,simenel2014b}, neutron droplets in harmonic potentials \cite{pudliner1996,bogner2011,gandolfi2011,maris2013,potter2014,shen2018,shen2019,zhao2020}, as well as exactly solvable models such as the Lipkin model \cite{lipkin1965} (see, e.g., \cite{severyukhin2006}).
For instance, the equation of state of infinite nuclear matter is a key ingredient to fitting protocols of energy density functional \cite{bender2003} used in nuclear density functional theory \cite{colo2020}.
Similar strategies are used in condensed matter physics, where the uniform electron gas (or jellium) is used in  the local-density approximation within density-functional theory (see, e.g., \cite{loos2016}).

Many properties, however, are not captured by homogeneous systems due to their translational invariance. 
Neutron drops, for instance, contain additional information on shell closures \cite{shen2018}, spin-orbit splitting \cite{maris2013,zhao2020} and surface properties \cite{potter2014,zhao2016} of nuclear systems. 
Neutron drops thus provide interesting platforms to compare results from ``ab initio'' approaches and density functional theories. 
These idealised finite systems are then complementary to homogeneous infinite matter. 

Quantum rings \cite{viefers2004} and their generalisation to spheriums \cite{loos2009} 
are another example of idealised systems which can provide valuable insights into the physics of complex interacting many-body systems. 
A continuum quantum ring is a system of particle wave-functions constrained to be on a ring of radius $R$. 
Experimentally, such systems interacting via the Coulomb interaction can be realised, e.g.,  in circular storage rings with heavy ion beams \cite{danared2002,steck1996}, and can be studied with Coulomb string models \cite{hasse2003}. 
One attractive aspect of quantum rings is that they can often be solved exactly, 
providing a benchmark to test various approximations to the quantum many-body problem.
Although many theoretical investigations of continuum quantum rings involve electron systems  
\cite{emperador2001,emperador2003,zhu2003,viefers2004,fogler2005b,aichinger2006,gylfadottir2006,loos2009,rasanen2009,manninen2009,loos2012,loos2013,rogers2017},
applications to atomic systems where interactions are typically short-range have also been considered, in particular to describe bosons in annular traps \cite{bargi2010,kaminishi2011,manninen2012,chen2019}.
To our knowledge, applications of quantum rings to nuclear systems have not been considered so far.

In this work we investigate the behaviour of quantum many-fermion systems exhibiting qualitative features of nucleons, such as finite range attraction with a repulsive core. 
Our  goal is to demonstrate  that basic features of nuclear systems, such as saturation and self binding, can be reproduced, paving the way for future applications of quantum rings to systems of nucleons with realistic nuclear interactions.
As a first application, exact (numerical) and mean-field (analytical) solutions with Yukawa type potentials are found and compared with $1/r$ Coulomb interaction. 

The paper is organised as follows. Quantum Monte-Carlo techniques are described in section~\ref{sec:MC}.
The quantum ring model and its analytical mean-field solutions are presented in section~\ref{sec:ring}. 
Results obtained by varying the number of particles and the ring radius $R$  are discussed in section~\ref{sec:results}. Conclusions from this study are drawn in section~\ref{sec:conclusion}.

\section{Quantum Monte-Carlo techniques}\label{sec:MC}

	Quantum Monte Carlo (QMC) approaches exploit the advances of modern parallel computing in the treatment of
	many body quantum systems.
	A weighted random exploration of the full configuration space is conducted,
	as opposed to the entire space, 
	where the independence of each configuration from the last allows for high parallelism.
	The result being that computational tractability is maintained far beyond the particle number limitations of
	other techniques which explicitly account for correlation. 
	A primary drawback however, is that the method is
	highly dependant on the form of trial wavefunction used, and as such requires significant knowledge of this
	form before any sensible optimisation can be conducted.
	
	QMC methods fall into two major categories: variational Monte Carlo (VMC) and 
	diffusion Monte Carlo (DMC). 
	In VMC the multivariate integration is performed by a weighted sum over random configurations with
	probability distribution $|\Psi|^2$ known as the Metropolis algorithm 
	\cite{1953JChemPhysMetropolis}.
	Should the wavefunction be exact then any configuration within this form will be of the
	same local energy. 
	This principle allows VMC computation of said energy to be repeatedly preformed with
	differing parameters in the trial wavefunction, seeking to minimise the variance or even energy directly.

	In contrast, 
	DMC takes a set of entire configurations and propagates them in imaginary time. The initial set of
	configurations is generated by a VMC computation. Past a sufficient period for equilibration, a
	statistical average is taken across all included configurations that is guaranteed as an upper bound on the
	true result given the nodal surface of the trial wavefunction is exact.
Using a wrong nodal structure would lead to a ``fixed-node'' systematic error. 
However, the nodal structure of the HF solution is exact for a Coulomb potential in 1D, making DMC calculations particularly suited as they are free of fixed-node error in such systems \cite{lee2011}. 
For finite range interactions such as Yukawa potentials, the nodal structure of 1D systems is expected to be the  same as in the case of the Coulomb interaction. 

	While in theory, imaginary time propagation will grant the desired result regardless of the initial input,
	in practise, a well optimised trial wavefunction is needed for reasonable efficiency.
	Under such circumstances DMC has proven to be as accurate or better than correlated wavefunction techniques
	used in quantum chemistry while remaining applicable to very large systems.
Here, both VMC and DMC calculations in the quasi-1D ring configuration are performed with the CASINO code \cite{needs2009,2019CASINOManual}. 
Similar calculations for a linear 1D system of electrons were performed by Lee and Drummond \cite{lee2011}.

\section{Quantum ring model}\label{sec:ring}

\subsection{Hamiltonian}

	We consider the problem of $n$ particles of alike spin 
	on a ring of radius $R$ with the interaction $V(r_{ij})$.
	For the particle $i$, its position on the ring is $x_i\in[0,\;2\pi R)$, and its signed `through-the-ring'
	distance to particle $j$ is
	\begin{alignat}{3}
		r_{ij}&=2R\sin\left(\frac{x_i-x_j}{2R}\right)\; .
	\end{alignat}
	The Seitz radius is given by
	\begin{alignat}{3}
		r_s=\frac{\pi R}{n}=\frac{1}{2\bar\rho}\; ,
	\end{alignat}
	where $\bar\rho=\frac{1}{2\pi R}\int_0^{2\pi R} dx \,\rho(x)$ is the average of the linear density $\rho(x)$. 
	Each particle occupies, in average, a section $2r_s$ of the ring. 

The Hamiltonian is defined by
	\begin{alignat}{3}
		H=-\frac{1}{2}\sum_{i=1}^n\frac{\partial^2}{\partial x_i^2}+\sum_{i<j}^nV(r_{ij})\;.
	\end{alignat}

\subsection{Variational space}

To describe the correlations induced by the interaction, we follow a method introduced by Jastrow \cite{jastrow1955}. 
	We seek a solution of the form
	\begin{alignat}{3}
		\label{wavefunction}
		\Psi^n&=\exp(J)\;\Psi_0^n\;,
	\end{alignat}
	where $\Psi_0^n$ is the Slater determinant
	\begin{alignat}{3}
		\label{slater}
		\Psi_0^n\propto\det\left[\exp(i m_i x_i/R)\right]\; ,
	\end{alignat}
	for
	\begin{alignat}{3}
		\nonumber
		m_i&\in\left\{-\frac{n-1}{2},-\frac{n-3}{2},\;\ldots\;,+\frac{n-3}{2},+\frac{n-1}{2}\right\}\;,\\
		i&\in\left\{1,\;\ldots,\;n\right\}\; .
	\end{alignat}

	 The Jastrow function $J$ is expressed as \cite{2004PRBNeeds}
	\begin{alignat}{3}
		J&=\sum_{i<j}\left[(|r_{ij}|-L_c)^3\;\Theta\left(L_c-|r_{ij}|\right)\sum_{k=0}^5\alpha_k
		|r_{ij}|^k\right]\;,
	\end{alignat}
	where $\Theta$ is the Heaviside step function and $L_c$ is a cutoff length.
In 1-dimensional systems, the Kato cusp
	condition at $r_{ij}=0$ \cite{1957CommPureApplKato} is satisfied 
with	
\begin{alignat}{3}
		\label{alpha1}
		\alpha_1&=\frac{3\alpha_0}{L_c}-\frac{\gamma}{(L_c)^3}\; ,
	\end{alignat}
	and  $\gamma=1/2$~\cite{loos2012}.
	The remaining $\alpha_k$ we seek to determine via VMC energy minimisation with the CASINO code \cite{2019CASINOManual}.
	This form of Jastrow is a complete power series up to order 8 in $r_{ij}$ with continuous 
	 first and second derivatives at $r_{ij}=L_c$.
	 We choose $L_c=\pi R$ which is greater than the ring diameter and thus allows for long-range correlations through the entire ring. 

	Using the wavefunction \eqref{wavefunction} we evaluate its corresponding energy per particle 
	\begin{alignat}{3}
		\label{epsilonn}
		\epsilon(n)=\frac{1}{n}\frac{\int(\Psi^n)^* H \Psi^n\;\mathrm{d}^n x_i}
						 {\int\big|\Psi^n\big|^2\;\mathrm{d}^n x_i } \;,
	\end{alignat}
	via VMC, or for the highest precision, DMC methods.

\subsection{Hartree-Fock  solution}

Restricting the variational space to a single Slater determinant corresponds to the Hartree-Fock mean-field approximation. 
	Interestingly, it can be shown \cite{2006PRLMitas,loos2013} 
	that the form \eqref{slater} of the Slater determinant is equivalent to the Vandermonde
	determinant 
	\begin{alignat}{3}
		\Psi_0^n&\propto\prod_{i<j}^n r_{ij}\; .
	\end{alignat}
\subsubsection{Kinetic energy}
	Using this form we can evaluate analytically the kinetic energy (KE) component of
	\eqref{epsilonn} as
	\begin{alignat}{3}
		\label{KEanalytic}
		\epsilon^{\mathrm{KE}}_{\mathrm{HF}}(n)&=\frac{n^2-1}{24R^2}\;.
	\end{alignat}
\subsubsection{Coulomb interaction}
	Similarly for the potential energy (PE) and $V(r_{ij})=1/|r_{ij}|$ we have \cite{loos2013}
	\begin{alignat}{3}
		\label{PEanalytic}
		\epsilon^{\mathrm{PE}}_{\mathrm{HF}}(n)&=
			\frac{2}{n\pi R}
			\sum_{i=0}^n \frac{1}{2i+1}
			\left(
				\begin{array}{c}
					n-i\\
					2\\
				\end{array}
			\right) \; ,
	\end{alignat}
	where $\left(\begin{array}{c}n\\r\\\end{array}\right)$ is the binomial coefficient.
	For $r_s=1$ and $n>20$, Eq.\ \eqref{PEanalytic} is well approximated by
	\begin{alignat}{3}
		\epsilon^{\mathrm{PE}}_{\mathrm{HF}}(n)\approx\frac{1}{2}\psi(n+3/2)-\frac{1}{2}\psi(1/2)-\frac{3}{4} \;,
	\end{alignat}
	where $\psi$ is the digamma function. 
	In the limit $n\to \infty$ 
	\begin{alignat}{3}
		\epsilon^{\mathrm{PE}}_{\mathrm{HF}}(n)\to\frac{1}{2} \psi(n)\;\;\mbox{or}\;\;\frac{1}{2}\ln(n)\;,
	\end{alignat}
	as both functions $\to\frac{1}{2}\sum_{k=1}^\infty\frac{1}{k}\;$.
\subsubsection{Yukawa interaction}
\begin{figure}[htbp]
		\includegraphics[width=1.0\columnwidth]{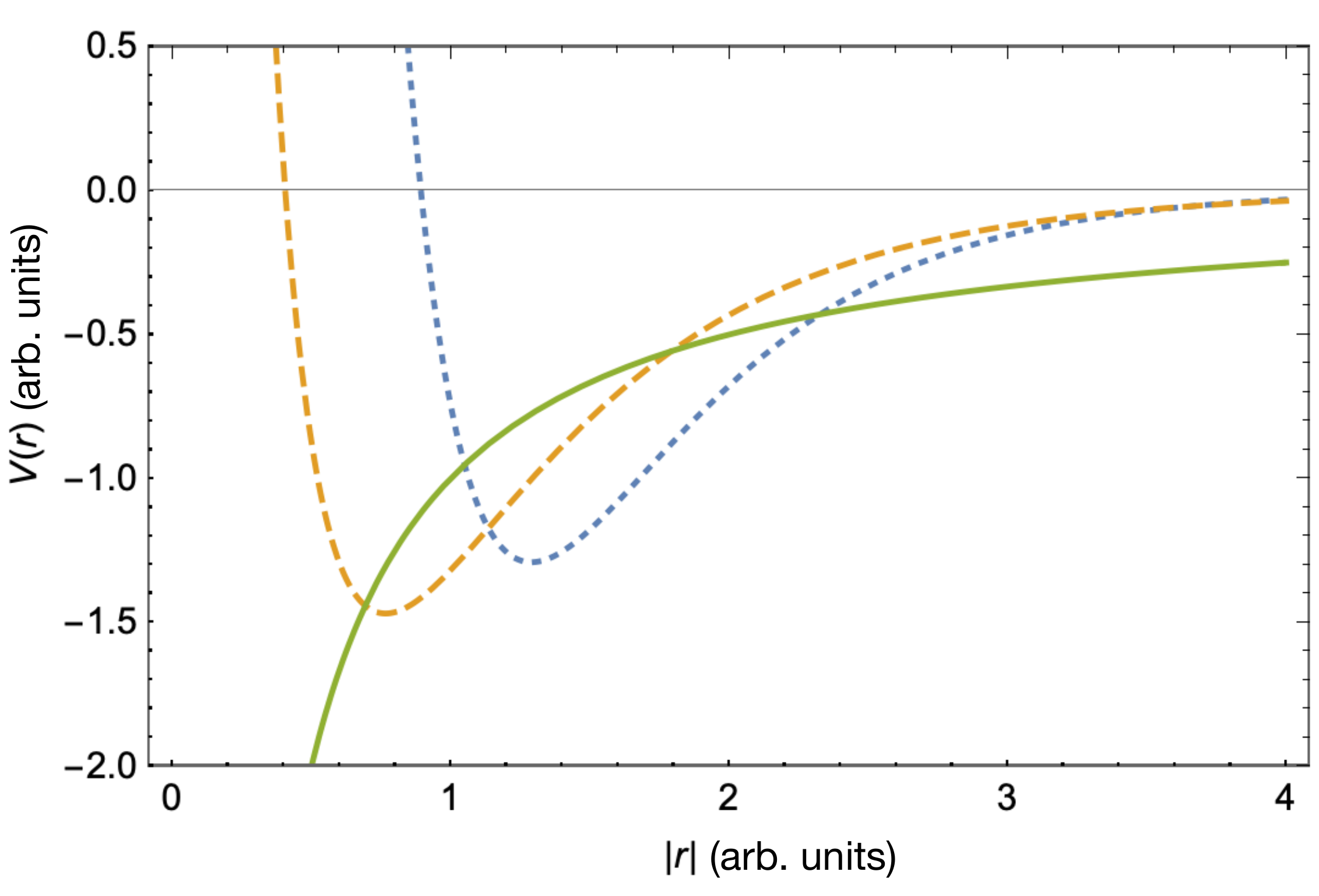}\\[0.2cm]
		\caption{
			Comparison between an attractive Coulomb potential (solid), and the two model nuclear
			potentials, Eq.\ \eqref{NucPot2} (short
			dashed) and Eq.\ \eqref{NucPot} (dashed).
			}
		\label{fig:potential}
	\end{figure}

	The remaining interactions  we consider are sums of attractive and repulsive Yukawa potentials to mimic 
	the interaction between nucleons. Two different parameterisations are considered:
	\begin{alignat}{3}
		\label{NucPot2}
		V^{\mathrm{nuc}}_1(r_{ij})&=\frac{1}{|r_{ij}|}\left[100e^{-2.0|r_{ij}|}
		-64e^{-1.5|r_{ij}|}\right]\; ,\\
		\label{NucPot}
		V^{\mathrm{nuc}}_2(r_{ij})&=\frac{1}{|r_{ij}|}\left[\phantom{1}12e^{-2.0|r_{ij}|}
		-\phantom{6}8e^{-1.0|r_{ij}|}\right]\; .
	\end{alignat}
	These two potentials are contrasted against $-1/|r_{ij}|$ in Fig.\ \ref{fig:potential}.

	For a Yukawa interaction of screening parameter $a$ we find the 
	Hartree-Fock potential energy to be given by
	\begin{alignat}{3}
		\label{PEYukanalytic}
		\epsilon^{\mathrm{PE}}_{\mathrm{HF,Y}}(n)&=
			\frac{1}{2n\pi R}
			\sum_{i=0}^n g(i,aR)
			\left(
				\begin{array}{c}
					n-i\\
					2\\
				\end{array}
			\right) \; ,
	\end{alignat}
	where
	\begin{alignat}{3}
		\label{PEYukg}
		g(i,aR)=2\sqrt{\pi}\sum_{j=0}^i (-1)^j
				\left(
				\begin{array}{c}
					2i+1\\
					2j+1\\
				\end{array}
				\right)
				\Gamma \!\left(i-j+1/2\right)\nonumber\\
				\times\Big[\Gamma \!\left(j+1\right) \,
				_1\tilde{F}_2\left(j+1;1/2,i+3/2;a^2R^2\right)\phantom{\Big]\; ,}\nonumber\\
			-aR \;\Gamma \!\left(j+3/2\right) \, _1\tilde{F}_2\left(j+3/2;3/2,i+2;a^2R^2\right)\Big]\; ,
	\end{alignat}
	$\Gamma$ is the gamma function, and $_p\tilde{F}_q$ is the regularised generalised hypergeometric
	function. Indeed Eq.\ \eqref{PEYukg} reduces to $4/(2i+1)$ for $a=0$, consistent with Eq.\
	\eqref{PEanalytic}, and goes to $0$ as $a\to\infty$.
\section{Results}\label{sec:results}
\subsection{Constant density}
\begin{figure}[htbp]
		\includegraphics[width=1.0\columnwidth]{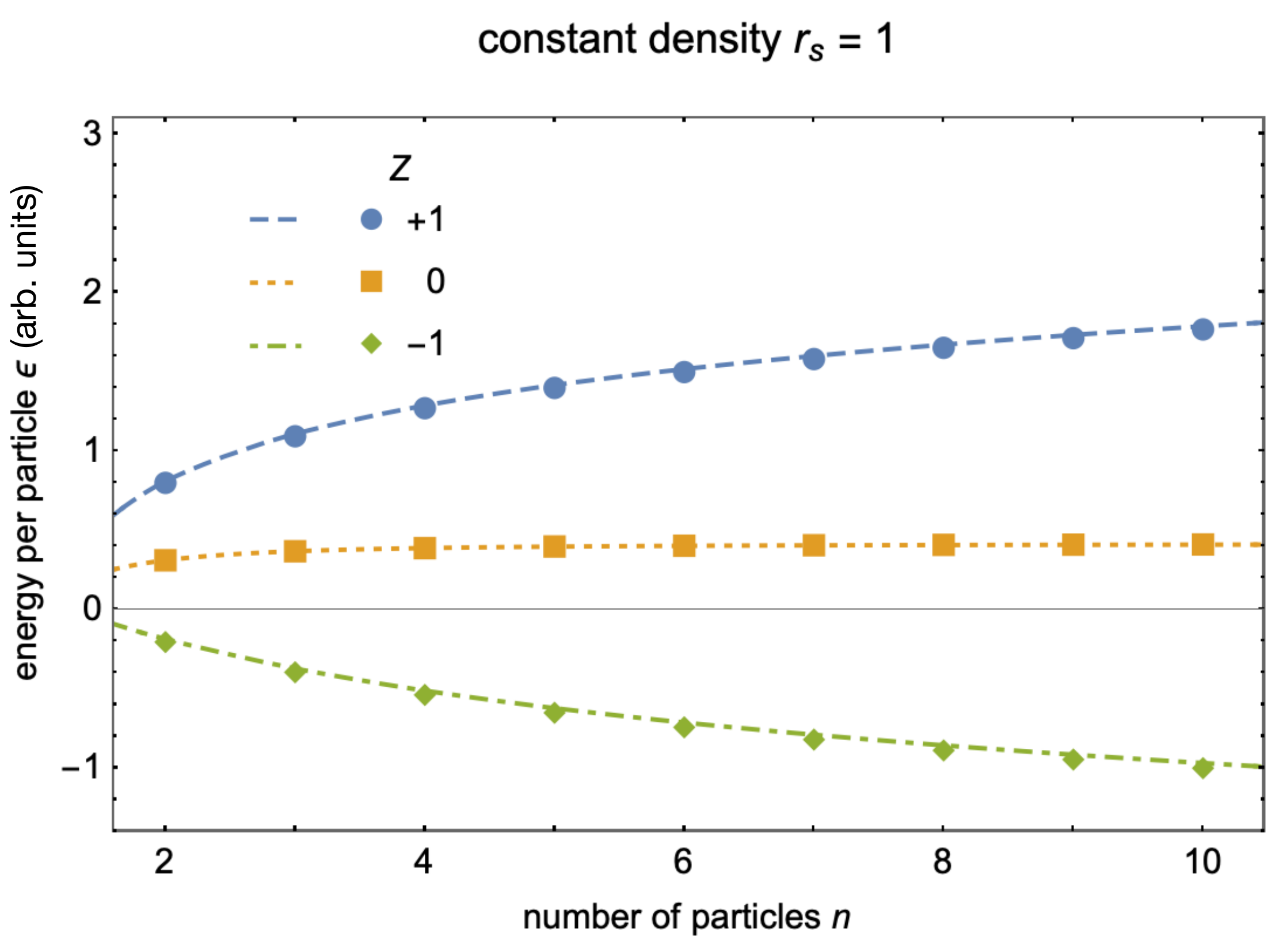}
		\caption{
			QMC (symbols) and HF (lines) energy per particle $\epsilon$ for $n$ alike spin particles 
			on a $r_s=1$ ring with an interaction
			potential $Z/r$ for $Z\in\{-1,0,+1\}$. 
			}
		\label{ZConstantRho}
	\end{figure}

	We begin our investigation taking the Seitz radius $r_s=1$ for attractive ($Z=-1$), zero ($Z=0$), and repulsive ($Z=+1$)
	Coulomb potentials $V(r_{ij})=Z/|r_{ij}|$.
	Figure\ \ref{ZConstantRho} shows the resulting energy  per particle from QMC calculations (symbols) and from the Hartree-Fock approximations (lines). 
	Due to the infinite range nature of the interaction the potential energy contribution continues to 
	increase/decrease as the system expands at constant density. 
	The zero case, which is comprised of purely the kinetic component, instead
	asymptotes to the value $\pi^2/24$.
	This case is found to match exactly the analytic
	expression Eq.\ \eqref{KEanalytic} as no optimisation of the wavefunction occurs. 
	In the interacting cases, the
	optimisation yields a lower energy than analytic expression 
	$\epsilon^{\mathrm{KE}}_{\mathrm{HF}}(n)+Z\epsilon^{\mathrm{PE}}_{\mathrm{HF}}(n)$
	from Eqs~\eqref{KEanalytic} and \eqref{PEanalytic} 
	by a few percentage points, as was similarly found for the repulsive case
	\cite{loos2013}.

\begin{figure}[htbp]
		\includegraphics[width=1.0\columnwidth]{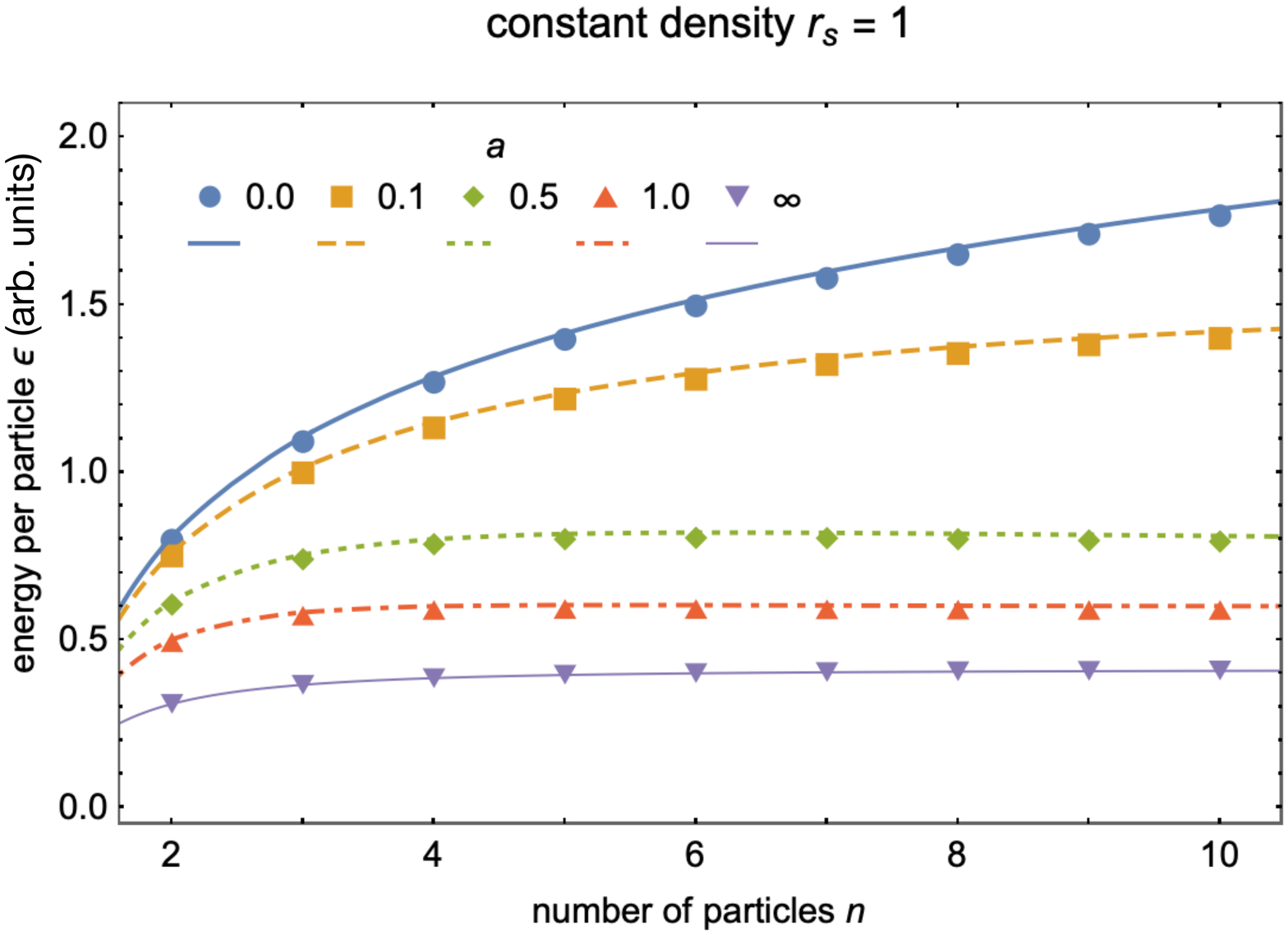}
		\caption{
			QMC (symbols) and HF (lines)  energy per particle $\epsilon$ for $n$ alike spin particles on a $r_s=1$ ring 
			with a repulsive Yukawa interaction
			potential $\exp(-ar)/r$ for $a\in\{0.0,0.1,0.5,1.0,\infty\}$. 
			}
			\label{aConstantRho}
	\end{figure}

	In Fig.\ \ref{aConstantRho} we instead consider Yukawa potentials $V(r_{ij})=\exp(-a|r_{ij}|)/|r_{ij}|$ with various screening parameters $a$.
	Introducing a finite range to the interaction causes the energy per particle to approach a constant value with increasing
	system size for a fixed density.
	Here the HF analytic expressions 
	are again found to overestimate the energy by several percent after Jastrow optimisation, but these
	differences approach zero as $a\to\infty$.
	This result is unsurprising as the potential component similarly goes to zero.
\subsection{Constant particle number}
	Having established the accuracy of the HF expressions for the individual Coulomb and Yukawa interactions we
	additionally consider the model nuclear interactions $V_{1,2}^{\mathrm{nuc}}$ in Eqs.~\eqref{NucPot2} and~\eqref{NucPot}. As can be seen in
	Fig.\ \ref{fig:potential}, the primary difference between these potentials is the distance at which their
	minima occurs, which is larger for $V^{\mathrm{nuc}}_1$ (short dashed line in Fig.~\ref{fig:potential}) due to a higher repulsion term simulating the hard-core part of the nuclear interaction. 
	
\begin{figure}[htbp]
		\includegraphics[width=1.0\columnwidth]{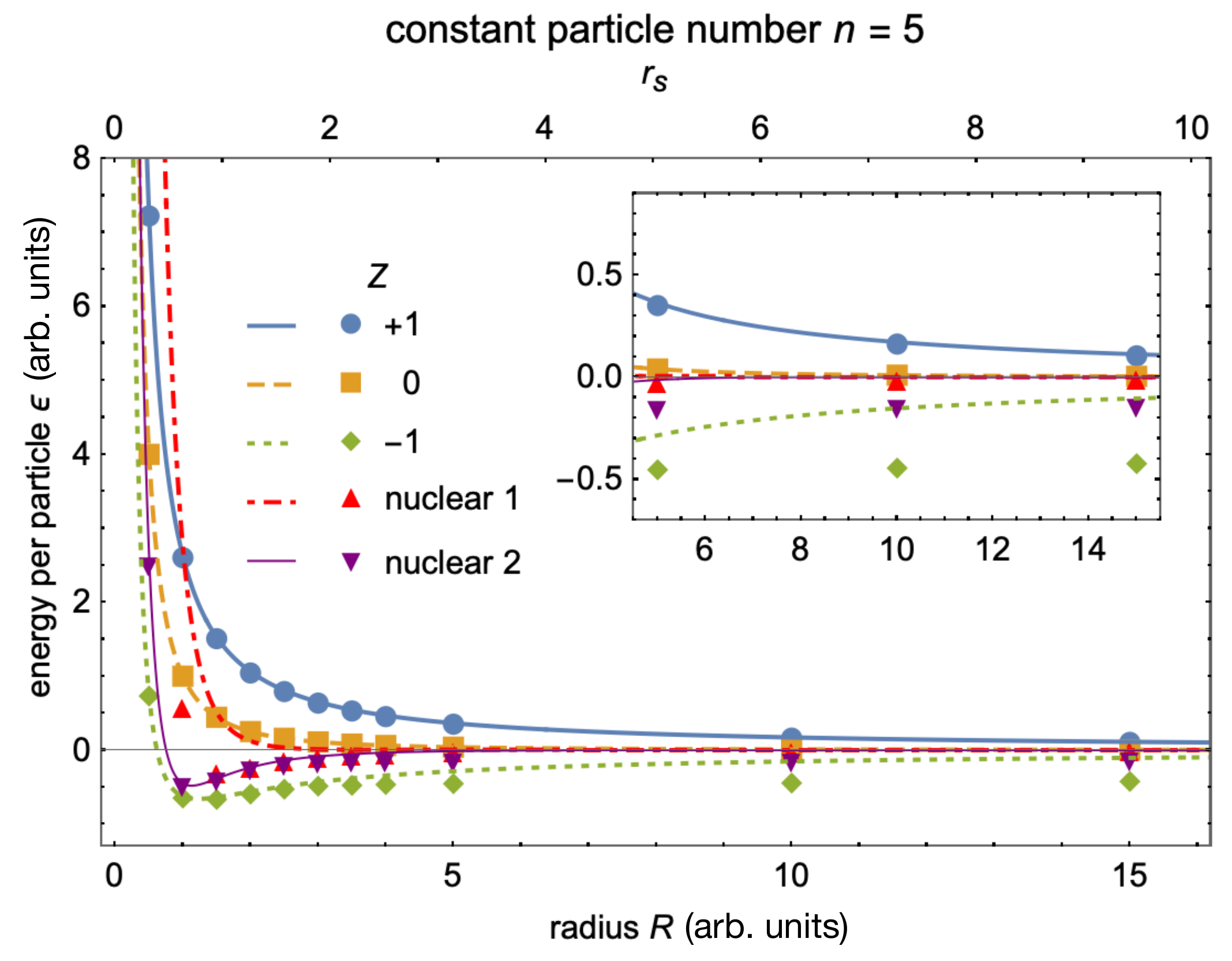}
		\caption{
			The energy per particle $\epsilon$ for 5 alike spin particles on a ring of varying radii
			with an interaction
			potential $Z/|r_{ij}|$ for $Z\in\{-1,0,+1\}$. 
			The two model nuclear interactions 
			in Eqs.~\eqref{NucPot2} and \eqref{NucPot}
			are depicted with
			triangles, right and inverted respectively.
			Each set of points is given alongside its corresponding 
			HF analytic expression from 
			Eqs.\ \eqref{KEanalytic}, \eqref{PEanalytic}, and \eqref{PEYukanalytic}.
The corresponding Seitz radii $r_s$ are given in the top axis. 			}
			\label{ZConstantNExtended}
	\end{figure}

	We contrast each interaction potential for a system
	of $n=5$ particles and varying ring radius $R$ in Fig.\ \ref{ZConstantNExtended}.
	For small radii the kinetic energy diverges as $1/R^2$ and, as such, we find the energy to become strongly
	positive in this region for all interactions. However, with increasing $R$, each interaction begins to exhibit their own 
	distinctive dependencies.
	The  Coulomb ($Z=1$) repulsion is still present for large $R$ due to its long-range $1/r$ dependency. In this case, the HF and exact results are in excellent agreement, indicating no or little beyond HF correlations.
	However, in the case of the potentials with an attractive component ($Z=-1$ and $V_{1,2}^{\mathrm{nuc}}$), the QMC results significantly deviate from the HF analytical results at large radii. 
	As illustrated by the inset of Fig.~\ref{ZConstantNExtended}, these systems are bound by a negative energy per particle which becomes constant at large $R$. 
	In this regime, a lower
	energy state is accessible from the particles preferentially bunching together compared to the base uniform
	case. An equilibrium distance between each particle is reached and further increases to the
	ring size no longer affect the energy.
	This behaviour is similar to the density saturation mechanism in nuclei. 
	The inability of the HF analytical expressions to reproduce this behaviour is due to the   space invariance of the mean-field Hamiltonian. As is well known in nuclear physics, however, such correlations can be incorporated at the mean-field level simply by breaking the symmetry under space translation~\cite{vautherin1972}.

	The difference between the HF predictions for the two nuclear-like potentials for $R<5$~a.u. deserves further discussions. 
	QMC predictions with $V_2^{\mathrm{nuc}}$ behave almost identically to the HF
	analytic expression. For $V_1^{\mathrm{nuc}}$, however, the QMC result is significantly below the HF case. 
	In fact,
	this latter analytic expression suggests there is no radius at which the system will be bound, whereas in
	contrast, the Jastrow optimised values behave similarly to $V_2^{\mathrm{nuc}}$ with negative energy above $R\sim 1$. 
	The reason behind this is the very strong repulsive nature of $V_1^{\mathrm{nuc}}$ for comparatively large
	$|r|$. In the HF state $\Psi_0^n$ the wavefunction is distributed irrespective of the interparticle interaction, which
	for $V_1^{\mathrm{nuc}}$, begins to significantly occupy regions of strong repulsion. Instead, for the correlated state $\Psi^n$,
	the Jastrow optimisation allows the wavefunction to redistribute spatially, producing fluctuations that lead to  significantly more localised  wave-functions minimising the  repulsion between the particles. 
	
\subsection{Constant radius}
\begin{figure}[htbp]
		\includegraphics[width=1.0\columnwidth]{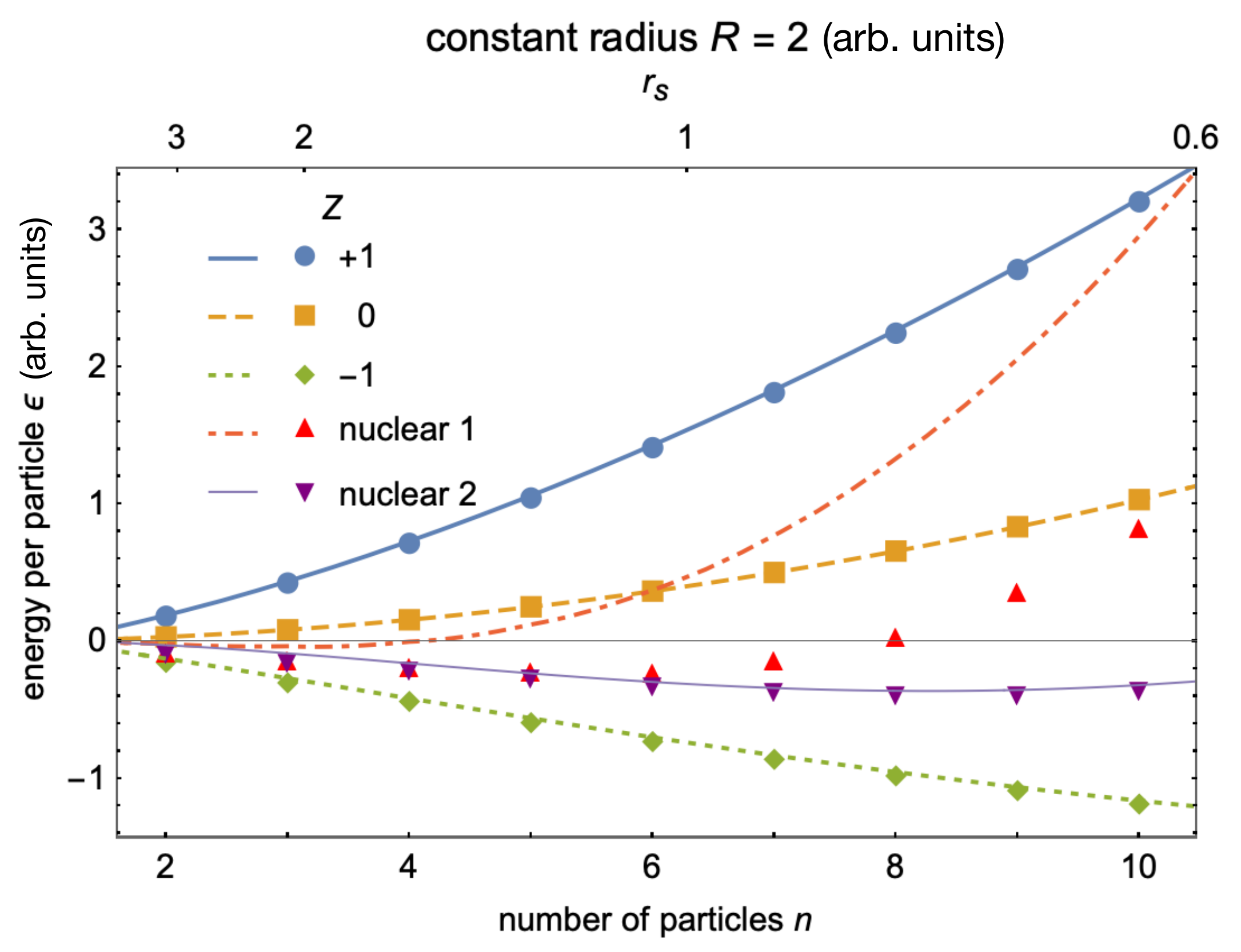}
		\caption{
Same as fig.~\ref{ZConstantNExtended} for $n$ alike spin particles on a ring of radius $R=2$~a.u.. 
}
\label{ZandNucConstantR}
	\end{figure}

	We now investigate  two regimes: small radii, where there is strong dependence on the ring size; and
	large radii, where the system becomes size independent for attractive interactions. The  
	energies for
	various particle numbers $n$ on a $R=2$ a.u. ring
	are given in Fig.\ \ref{ZandNucConstantR}. In the $Z=\pm1$ cases,
	the QMC results are again slightly below the analytic HF expressions, indicating small correlations. 
As shown by the $Z=0$ case, the energy of non-interacting particles increases with $n$ due to the Pauli exclusion principle. 
For $V_{1,2}^{\mathrm{nuc}}$, the systems first gain energy with increasing $n$ due to the attractive nature of the interaction. At large $n$, however, they become less bound (and eventually unbound) due to the Pauli induced increase of the kinetic energy and, in particular for $V_1^\mathrm{nuc}$, the strong short-range repulsion. 
However, as was observed in Fig.\ \ref{ZConstantNExtended},
	there is significant deviation between HF and QMC predictions for $V_1^{\mathrm{nuc}}$.
	 The Jastrow optimisation in QMC leads to
	far lower energies than available for the $\Psi^n_0$ (HF) form of the wavefunction. 
	Again, this is due to the
	redistribution of the $\Psi^n$ correlated wavefunction to minimize the strong repulsion.
	\begin{figure}[htbp]
		\includegraphics[width=1.0\columnwidth]{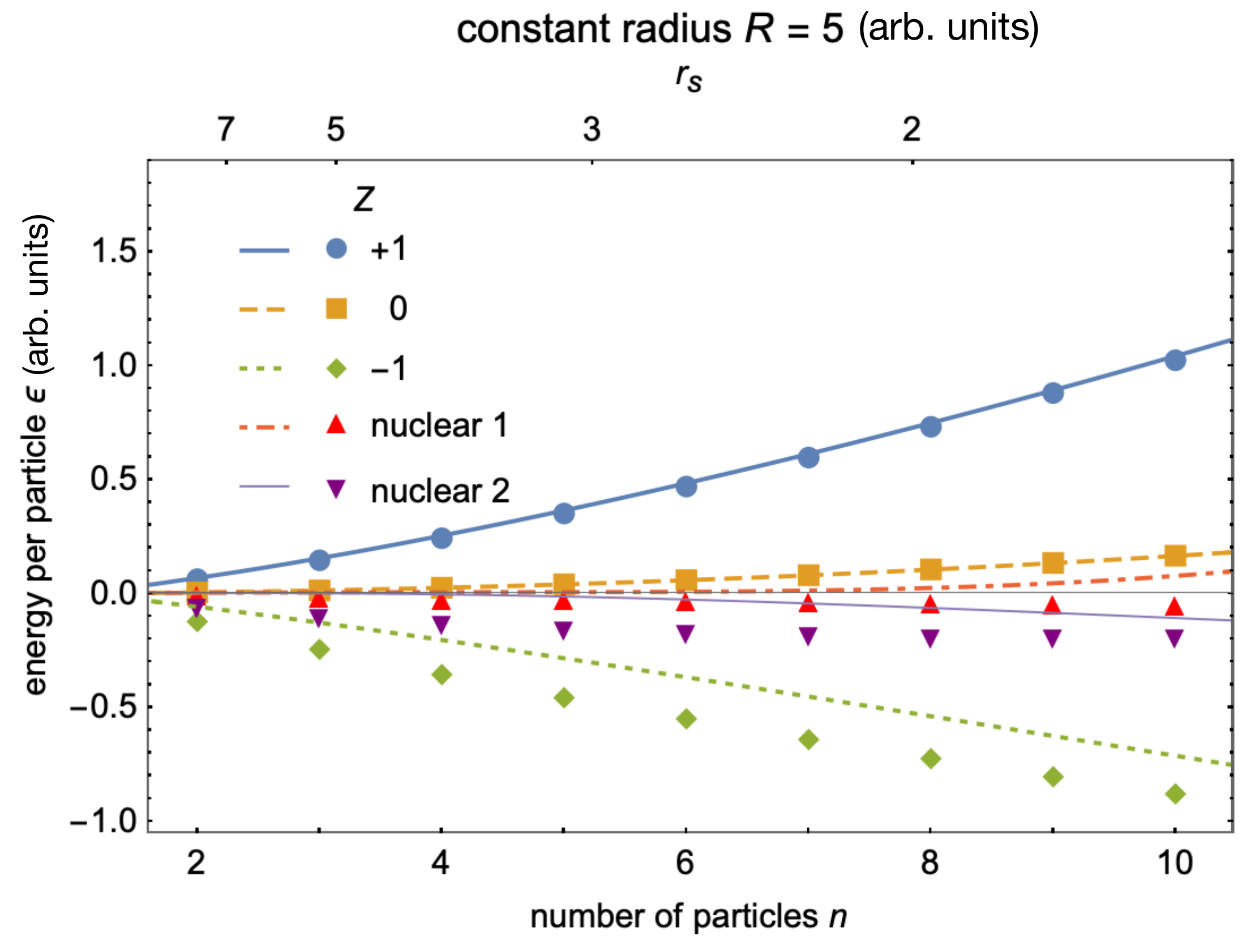}
		\caption{
Same as fig.~\ref{ZandNucConstantR} for ring of radius $R=5$~a.u..}
			\label{ZandNucConstantR5}
	\end{figure}

	Figure\ \ref{ZandNucConstantR5} presents the energy per particle as a function of particle number $n$ 
	for $R=5$ a.u.. 	
	The repulsive Coulomb interaction ($Z=1$) and  non-interacting  ($Z=0$) cases behave as before. 
	For the three attractive interactions ($Z=-1$ and $V^\mathrm{nuc}_{1,2}$), however, 
	there is a large disagreement between the QMC and HF results. 
	Significant more binding is predicted in QMC calculations, indicating spatial correlations not captured in the HF method. 
	These correlations are due to the attraction which make the particles collate together in a region of space, forming a self-bound system. 
These systems would eventually become unbound if we keep increasing the number of particles, or equivalently decrease the Seitz radius $r_s$, as the Pauli induced kinetic energy would also increase.

Effects of the interaction range are also visible on Fig.~\ref{ZandNucConstantR5}.
With a  long range attraction ($Z=-1$), the energy keeps decreasing with the number of particles as each additional particle is able to interact with all the other particles of the localised system. 
With nuclear like short range interactions, however, a saturation of the binding energy per particle is observed as each additional particle is only able to interact with its closest neighbours. 
This saturation behaviour is typical of short range interactions as observed in nuclear systems. 

\section{Conclusions}\label{sec:conclusion}

The quantum ring model has been applied to systems of fermions interacting with different types of potentials, both attractive and repulsive, and with various ranges. 
In particular, interactions sharing qualitative features of the nuclear interaction between nucleons, i.e., a short-range repulsion and a finite range attraction, have been considered. 

Thanks to the ring geometry, exact numerical solutions were obtained with the Quantum Monte-Carlo technique. 
Analytical solutions of the Hartree-Fock self-consistent mean-field equations were also found. 
Comparisons between HF and QMC predictions give an insight into the role of spatial correlations in the many-body systems. 
In many cases, the HF method agrees well with the QMC results. 
However, a strong short-range repulsion induces correlations to minimise the repulsion at high density.  
Similarly, spatial correlations are found in large rings, allowing for the formation of localised self-bound systems which, in the case of nuclear-like interactions, have constant binding energy per particle as in nuclei. 
Such correlations could be introduced at the mean-field level using the standard technique of breaking symmetries of the HF Hamiltonian. 

These shared features with nuclear systems encourage the application of quantum rings and their generalisations to higher dimensions (hypersphere)  \cite{loos2009}
with realistic interactions between nucleons to test approximations to the nuclear many-body problem, as well as in the construction of energy density functionals.

The study of the behaviour of correlations with the system's dimensionality is interesting in its own right. 
For repulsive electrons, the magnitude of correlation (i.e. beyond HF) effects is profoundly dependent on the number of dimensions. Specifically, correlation effects are moderate in 1D  (e.g., in linear \cite{schulz1993,fogler2005a} or  ringium \cite{rogers2017} configurations), much stronger in 2D (e.g. spherium), moderate in 3D (e.g. glomium) \cite{loos2009}, and then monotonically weaker in higher dimensions. 
For attractive potentials, however, we would expect strong correlations (responsible for the formation of self-bound systems) to persist with increasing number of dimensions as long as the particles have the possibility to remain ``near'' to one-another. 
Nevertheless, the quantitative evolution of the magnitude of these correlations in attractive systems with dimensionality remains to be studied. 

Quantum hyperspheres could  play a complementary role as other idealised systems in nuclear physics such as neutron drops and infinite nuclear matter. 
It is also possible that such simplified geometries could find applications to describe semi-realistic systems, such as neutron skins and haloes in exotic nuclei. 

\acknowledgements
CS thanks P. Gill for introducing the quantum ring model, for useful discussions at the early stage of this work, as well as feedback on the manuscript. 
We thank P.-F. Loos for literature advice, for his suggestion of using the CASINO code, and feedback on the manuscript. 
Useful discussions with H. Witek are also acknowledged. 
CS is grateful to J. Nigam for her analytical investigation of the quantum ring model with Yukawa interactions prior to this work. 
We thank M. Towler for giving us permission  to make modifications to the CASINO code.  
This work has been supported by the Australian Research Council Discovery Project (Projects No. DP180100497 and DP190100256) funding scheme.
This work was supported by computational resources provided by the Australian Government through the National Computational Infrastructure (NCI) under the ANU Merit Allocation Scheme.

\bibliography{VU_bibtex_master.bib}

\end{document}